\documentclass[twocolumn, a4paper, times, trackchanges]{aastex61}

\usepackage[english]{babel}
\usepackage[utf8x]{inputenc}
\usepackage[T1]{fontenc}

\usepackage{amsmath}
\usepackage{graphicx}
\usepackage[colorinlistoftodos]{todonotes}
\usepackage{natbib}

\begin{document}
\title{Evolution of the Solar Lyman-Alpha line profile during the solar cycle}

\correspondingauthor{Izabela Kowalska-Leszczynska}
\email{ikowalska@cbk.waw.pl}

\author{Izabela Kowalska-Leszczynska}
\affiliation{Space Research Centre, Polish Academy of Sciences,\\
Bartycka 18A, 00-716 Warsaw, Poland}

\author{Maciej Bzowski}
\affiliation{Space Research Centre, Polish Academy of Sciences,\\
Bartycka 18A, 00-716 Warsaw, Poland}

\author{Justyna M. Sokół}
\affiliation{Space Research Centre, Polish Academy of Sciences,\\
Bartycka 18A, 00-716 Warsaw, Poland}

\author{Marzena A. Kubiak}
\affiliation{Space Research Centre, Polish Academy of Sciences,\\
Bartycka 18A, 00-716 Warsaw, Poland}

\begin{abstract}
Recent studies of interstellar neutral (ISN) hydrogen observed by the Interstellar Boundary Explorer (IBEX) suggested that the present understanding of the radiation pressure acting on hydrogen atoms in the heliosphere should be revised. There is a significant discrepancy between theoretical predictions of the ISN H signal using the currently used model of the solar Lyman-$\alpha$ profile by \citet[TB09]{tarnopolski_bzowski:09} and the signal due to ISN H observed by IBEX-Lo. We developed a new model of evolution of the solar Lyman-$\alpha$ profile that takes into account all available observations of the full-disk solar Lyman-$\alpha$ profiles from SUMER/SOHO, provided by \citet[L15]{lemaire_etal:15a},  covering practically the entire 23rd solar cycle. The model has three components that reproduce different features of the profile. The main shape of the emission line that is produced in the chromosphere is modeled by the kappa function; the central reversal due to absorption in the transition region is modeled by the Gauss function; the spectral background is represented by the linear function. The coefficients of all those components are linear functions of the line-integrated full-disk Lyman-$\alpha$ irradiance, which is the only free parameter of the model. The new model features potentially important differences in comparison with the model by TB09, which was based on a limited set of observations. This change in the understanding of radiation pressure, especially during low solar activity, may significantly affect the interstellar H and D distributions in the inner heliosphere and their derivative populations.

\end{abstract}

\section{Introduction}
\label{sec:intro}
The solar resonance Lyman-$\alpha$ line is the brightest one in the EUV part of the solar spectrum. Full disk averaged profile of this line and its evolution during the cycle of solar activity is of high importance in various astrophysical contexts, from planetary and cometary atmospheres to the heliosphere. In the heliospheric context, which is the main topic of this paper, solar Lyman-$\alpha$ photons interact resonantly with neutral hydrogen (H) and deuterium (D) atoms. This interaction is responsible on one hand for the creation of the heliospheric backscatter glow, and on the other hand for the effect of solar radiation pressure \citep{wilson:60}, which modifies the trajectories of H and D atoms in the heliosphere \citep{axford:72, tinsley:71a}. Due to their very low density, neutral atoms in the heliosphere move without collisions with each other, governed by a net force due to solar gravity and radiation pressure. The effect of radiation pressure counteracts the solar gravity force. The magnitudes of the forces of radiation pressure and solar gravity force are comparable with each other. As a result, the spatial distribution of interstellar H inside the heliosphere, and thus the distribution of the heliospheric backscatter glow in the sky and of the fluxes of H$^+$ pickup ions (PUIs) are very susceptible to details of the solar output in the Lyman-$\alpha$ line. Since measurements of interstellar neutral (ISN) H, its derivative populations, and the heliospheric backscatter glow are very important sources of information on the processes in the heliosphere, accurate understanding of radiation pressure is of major importance for heliospheric studies.

It was discovered very early in the space age that the profile of the solar Lyman-$\alpha$ line features self-reversal \citep{purcell_tousey:60a, purcell_tousey:60c} and that the wavelength- and disk-integrated irradiance varies considerably with time \citep{blamont_vidal-madjar:71a}. The variation during the solar cycle is by a factor of 2. The irradiance is usually reported as a daily value, with variations due to the solar rotation ($\sim 27$~days), solar cycle ($\sim 11$~years) and even longer, with the amplitude of monthly variation of $\sim 5$\% during low solar activity, up to $\sim 20$\% during high activity \citep{woods_etal:05a}. 

The Lyman-$\alpha$ line is created within the chromosphere (wings of the profile), but the central core with the self-reversal is due to the absorption in the lower corona \citep{avrett:92a}. This self-reversal is observed in most of the line profiles observed in the quiet-Sun regions \citep{tian_etal:09e}, but in sunspots and active regions the profiles are not self-reversed \citep{tian_etal:09c}, similarly as in the flares. The percentage of the solar disk covered with these features varies during the solar activity cycle, and the line- and disk-integrated irradiance is tightly correlated with the sunspot number and other solar activity indicators. Furthermore, the line features limb darkening with a related variation of the profile shape in the central reversal, which seems to be a function of the angle off the disk center \citep{bonnet_etal:78a}. Therefore it is not easy to predict the shape of disk-integrated profile. The active phenomena on the solar disk are responsible for a considerable fraction of the total irradiance and accordingly they may significantly contribute to the disk-integrated line profile. 
We may get different disk-integrated profiles for the same value of disk- and wavelength-integrated irradiance $I_{\mathrm{tot}}$, depending on how the solar disk was covered with sunspots, active regions etc.

First attempts to obtain a disk-averaged profile of the solar Lyman-$\alpha$ line was undertaken by \citet{bonnet_etal:78a} and \citet{lemaire_etal:78a} based on observations from the Earth satellite OSO-8. The self-reversed structure of the profile was obtained, but at the center of the profile a sharp absorption feature was found due to the absorption of solar radiation in the exosphere. This absorption hampered determining the exact level of spectral irradiance close to the center of the line because this requires a sufficiently accurate model of the H density distribution in the exosphere. This problem troubled also subsequent observations of this line. It was important for the heliospheric studies because the exospheric absorption is in the most important spectral band for the resonance radiation pressure acting on ISN H atoms inside the heliosphere. 

First absorption-free profiles were obtained from the SUMER instrument on board SOHO \citep{wilhelm_etal:95a}, which is located in the vicinity of the Sun-Earth L1 Lagrange point. The measurements were performed between 1995 and 2009, i.e., during an interval covering two minima (1995, 2008) and one maximum (2000--2002) of solar activity. They were published in a series of papers by \citet{lemaire_etal:98, lemaire_etal:02, lemaire_etal:05}, and L15. In our paper we use the data presented in the last of these papers.

In the classical approach, the solar radiation pressure force was considered as a constant force decreasing with the square of solar distance. This approximation is valid when the density of ISN H atoms inside the heliosphere is low enough to enable neglecting the absorption of solar Lyman-$\alpha$ photons. As a result of this approximation, the force of radiation pressure can be represented by a constant factor $\mu$ compensating the solar gravity force \citep{axford:72}. The effective force responsible for the motion of hydrogen atoms in the heliosphere was approximated by the formula
\begin{equation}
\vec{F}(\vec{r}, \mu) = -G\,M_{\sun}\, m_{\mathrm{H}}(1 - \mu) \frac{\vec{r}}{r^3},
\label{eq:simpleEqMotion}
\end{equation} 
where G is the solar gravity constant, $M_{\sun}$ the solar mass, $m_{\mathrm{H}}$ the mass of hydrogen atom, $\mu$ is the  dimensionless gravity compensation factor, and $\vec{r}$ the radius-vector. This approximation was used by several authors \citep[e.g.,][]{fahr:78, fahr:79, thomas:78, wu_judge:79a} to construct the so-called ``hot model'' of the density distribution of ISN H inside the heliosphere. ``Hot model'' takes into account the streaming and thermal spread of the H atoms entering the heliosphere and their ionization losses due to charge exchange with solar wind and photoionization. It also takes into account the modification of the atom kinematics due to radiation pressure, assuming that solar gravity is compensated by a constant factor $\mu$. 

However, it was very early realized that due to the self-reversed shape of the profile of the solar Lyman-$\alpha$ line the resonant radiation pressure acting on H and D atoms in the heliosphere must be a function of radial velocities of these atoms relative to the Sun \citep{axford:72}. Since the line-integrated irradiance in the solar Lyman-$\alpha$ line varies with time, at a given instant of time $t$ it will be a function of both solar distance and the heliographic longitude and latitude of the vantage point given by the radius vector $\vec{r}$. Effectively, the coefficient $\mu$ in Equation~\ref{eq:simpleEqMotion} becomes a function of time, location of the atom in space, and radial velocity of the atom: $\mu = \mu(t, \vec{r}, v_r)$. As a result, trajectories of H and D are not the Keplerian hyperbolae and must be calculated numerically. This was first shown by \citet{grzedzielski_sitarski:75a}, who compared hyperbolic trajectories with those obtained numerically for the case of $\mu = \mu(v_r)$ (i.e., radiation pressure depending solely on radial velocity of the atom). 

The first attempt to account for the dependence of radiation pressure on atom radial velocities in the modeling of ISN gas was published by \citet{tarnopolski_bzowski:08a} for D and TB09 for H. These authors extended the Warsaw Test Particle Model \citep[WTPM,][]{rucinski_bzowski:95a, bzowski_etal:97} of the evolution of density and bulk velocity of ISN H in the heliosphere to take this effect into account. TB09 approximated the shape of the solar Lyman-$\alpha$ line with an analytic function, which depended linearly on the instantaneous disk-integrated Lyman-$\alpha$ irradiance $I_{\mathrm{tot}}$. The basis of the model were nine profiles published by \citet{lemaire_etal:02}. TB09 was subsequently widely used in the studies of ISN H in the heliosphere \citep[e.g.,][]{bzowski_etal:08a, izmodenov_etal:13a, schwadron_etal:13a, fayock_etal:15a, katushkina_etal:15b} and of energetic neutral atoms in the heliosphere \citep[e.g.,][]{bzowski_tarnopolski:06a, bzowski:08a, bzowski_etal:13a, mccomas_etal:10c, mccomas_etal:12c, mccomas_etal:14b, mccomas_etal:17a, swaczyna_etal:16a}. It was also used in the pioneering studies of ISN D in the heliosphere, first modeling \citep{tarnopolski_bzowski:08a, kubiak_etal:13a}, and subsequently experimental, which resulted in the detecton of ISN D at the Earth's orbit by IBEX \citep{rodriguez_etal:13a, rodriguez_etal:14a}. 

However, \citet{schwadron_etal:13a} and \citet{katushkina_etal:15b} suggested that current models of the distribution of ISN H inside the heliosphere that are using the solar Lyman-$\alpha$ line profiles from TB09 seem to be unable to reproduce the ISN H flux sampled by IBEX-Lo. These authors suggested that a better agreement with observations could be obtained if a different profile of the solar Lyman-$\alpha$ line was adopted, with a larger horn-to-center ratio. 

Here, we construct a new model of evolution of the spectral profile of the solar Lyman-$\alpha$ line as a function of wavelength-integrated irradiance, based on all available observations of this profile, published by L15. The data used and their uncertainties are presented in Section~\ref{sec:data}. The models used to approximate the Lyman-$\alpha$ profile in the past are discussed in Section~\ref{sec:pastModels}. The model that we have developed along with fitting procedure is in Section~\ref{sec:modelIKL} and the results are shown in Section~\ref{sec:results}. We discuss the fidelity of our model in Section~\ref{sec:discussion}. Since the radial velocity-dependent radiation pressure is even more important for the distribution of ISN D than for ISN H, we adapt the obtained model to D in Section~\ref{sec:results:D}. We close by offering some concluding remarks in Section~\ref{sec:summary}. 

\section{Data}
\label{sec:data}

\subsection{Line profiles}
\label{sec:data:obsProfiles}

\begin{figure}
\centering
\includegraphics[width=1.0\columnwidth]{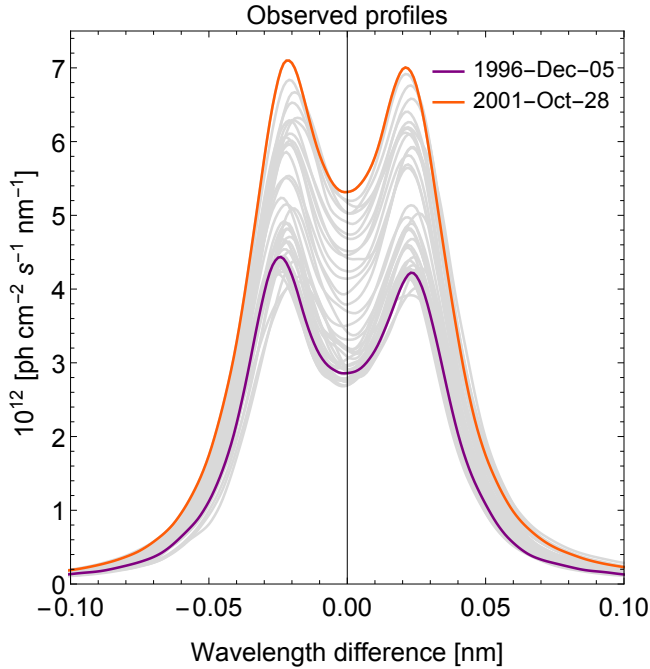}
\caption{
%{\em{fig:obsLemaire}} 
Observed profiles of disk-averaged solar Lyman-$\alpha$ line given by L15. Orange line represents a profile taken during maximum of solar activity (Oct 28th, 2001), while purple line shows a profile taken near the minimum of activity (Dec 5th, 1996). The horizontal scale is in nanometers off the central wavelength of the line at 121.567~nm.}
\label{fig:obsLemaire}
\end{figure}

The basis for our model are observations of disk-integrated profiles of the solar Lyman-$\alpha$ line measured by L15 using SUMER on board SOHO\footnote{Available from http://vizier.cfa.harvard.edu/viz-bin/VizieR?-source=J/A+A/581/A26}. These profiles are shown collectively in Figure~\ref{fig:obsLemaire}. There are 43 profiles observed between May 1996 and April 2009. The times of observations are illustrated in Figure~\ref{fig:totalFlux}, where the daily values of the wavelength-integrated irradiance for the observation days are superimposed on daily irradiance values from the composite Lyman-$\alpha$ series from LASP \citep{woods_etal:05a}. As evident, approximately half of the observations were taken during low solar activity, while a little less than a half during the maximum of solar activity in 2000--2003.

Details of the observations and processing applied to obtain the profiles scaled in the absolute units are given in L15 and in the earlier papers from those authors, therefore here we only point out some important aspects. The SUMER instrument had not been designed to provide disk-averaged profiles and therefore a specially devised observation scheme suggested by J.L.~Bertaux had to be used. Effectively, the instrument measured the profile of the solar Lyman-$\alpha$ radiation scattered inside the instrument optics when the boresight was directed $\sim 2000 \arcsec$ off the disk center. To safeguard a more or less uniform contribution from all parts of the solar disk, observations from several pointings around the solar disk were averaged. Since the angular distances of the boresight during a series of observations aimed at obtaining an individual profile for a given day were sometimes different, the spectral intensities from those individual pointings had to be rescaled to a common offset distance. The measurement process for one composite profile took more than four hours. 

The resulting profile is composed of 281 points at wavelengths equally spaced with a 0.001 nm pitch. During the observations, the profiles had been sampled at several wavelengths and the observations were subsequently deconvolved to eliminate the effect of the instrument point spread function. The final data product has a much higher resolution than the spectral resolution of the instrument and does not show any significant point-to-point scatter. The absolute scaling of the profiles was performed by the observers by requiring that the integral of the observed profile within the boundaries $\pm 0.15$~nm around the nominal wavelength of the line center is equal to the daily value of the composite series of the Lyman-$\alpha$ irradiance for a given day of observation (see Table A.1 from L15).
The uncertainties in the data set are complex. The uncertainty of individual profiles are given by the authors of the observations at 15\%. It seems, however, that an additional source of uncertainty is the systematic uncertainty of the LASP composite irradiance, which is also $\pm 15$\% \citep{woods_etal:05a}. This uncertainty results in systematic upward or downward offset of the entire data product, and it will result in a systematic uncertainty of the radiation pressure for H and D by the identical percentage. Additionally, because the integration time used to obtain the measurement of the LASP daily datum and that used during the spectral observations of the profiles were different, and because of the inevitable measurement scatter in the LASP observations, the absolute scaling of each of the individual profiles is biased relative to the other profiles from the sample, which results in random upward or downward shifts of the profiles relative to each other. Effectively, this uncertainty introduces correlations to the entire data set. The data points pertaining to individual profiles are stronger correlated than data points from different profiles, and absolute accuracies of different profiles are different. Therefore it is not reasonable to expect that the final model of the evolution of the profile during the solar cycle will fit to all of the observed profiles with the same accuracy.

\subsection{Total Lyman-$\alpha$ irradiance}
\label{sec:data:totalFlux}
The composite time series of the solar Lyman-$\alpha$ irradiance was recapitulated by \citet{bzowski_etal:13a}. The total irradiance, which we denote $I_{\mathrm{tot}}$, is based on direct daily observations, with occasional gaps
 filled with values derived from proxies. The absolute calibration is based on UARS/SOLSTICE\footnote{Available from ftp://laspftp.colorado.edu/pub/solstice/composite\_lya.dat}
 \citep{woods_etal:96a, woods_etal:00}.
Observations from later experiments, TIMED and SORCE, are scaled to this calibration. The irradiance is integrated over the 1~nm interval from 121 to 122~nm. The series cover the time interval since 1947 until
 present. The $I_{\mathrm{tot}}$ values for times before 1979 are obtained from scaling the solar F10.7 radio flux \citep{tapping:13a}. More recently, gaps have been filled mostly using the Mg~II core to wing
 index \citep{viereck_puga:99}. The use of proxies adds additional intermittent increase in the uncertainty of the daily irradiance values. 

\begin{figure}
\centering
\includegraphics[width=1.0\columnwidth]{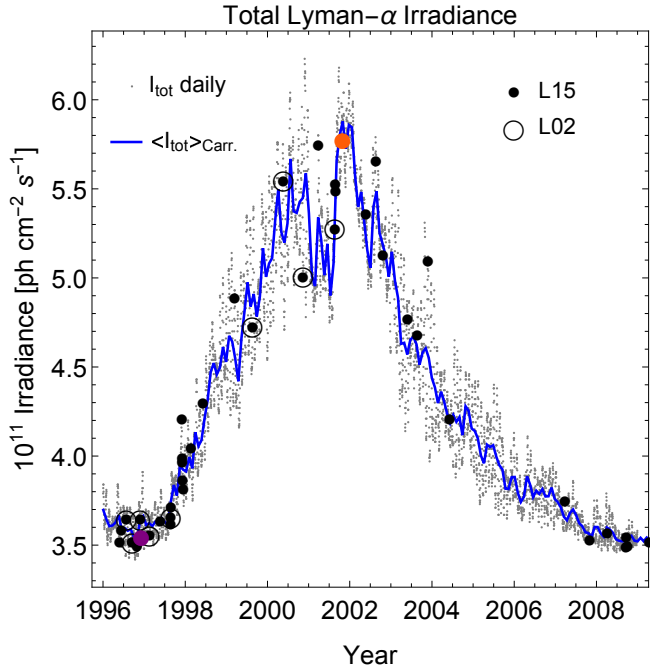}
\caption{
%{\em{fig:totalFlux}} 
Composite Lyman-$\alpha$ irradiance $I_{\mathrm{tot}}$ from LASP. The daily values are marked by gray dots, while the values averaged over Carrington period are marked by the blue line. The values corresponding to the days when the Lyman-$\alpha$ line profiles were observed by L15 are marked with the black filled circles, and those used to develop the TB09 model with the black unfilled circles (Observations denoted by L02 published by \citet{lemaire_etal:02}). Additionally we have marked two profiles (near solar minimum, taken on Dec 5th, 1996, denoted by the purple filled circle, and near solar maximum, taken on Oct 28th, 2001, denoted by orange filled circle), which are further discussed in Figures~\ref{fig:obsLemaire} and \ref{fig:fitExmp}.}
\label{fig:totalFlux}
\end{figure}

The values of the composite solar Lyman-$\alpha$ irradiance from LASP are listed at a daily cadence but they are not daily averages. Actually, the measurements are taken during a fraction of the day. Also the data used for proxies are not daily-average values, even though they are listed at a daily cadence. Therefore the daily $I_{\mathrm{tot}}$ values used by L15 to scale the Lyman-$\alpha$ profiles do not precisely correspond to the times of profile observations. 
The daily values of $I_{\mathrm{tot}}$ directly from the LASP are shown as gray dots in Figure~\ref{fig:totalFlux}, and the Carrington averages by the blue line. The daily values of $I_{\mathrm{tot}}$ for the dates corresponding to profiles observed by L15 are shown with black dots, and the subset from \citet{lemaire_etal:02} used by TB09, are taken in black circles.

\begin{figure*}
\centering
\includegraphics[width=0.35 \textwidth]{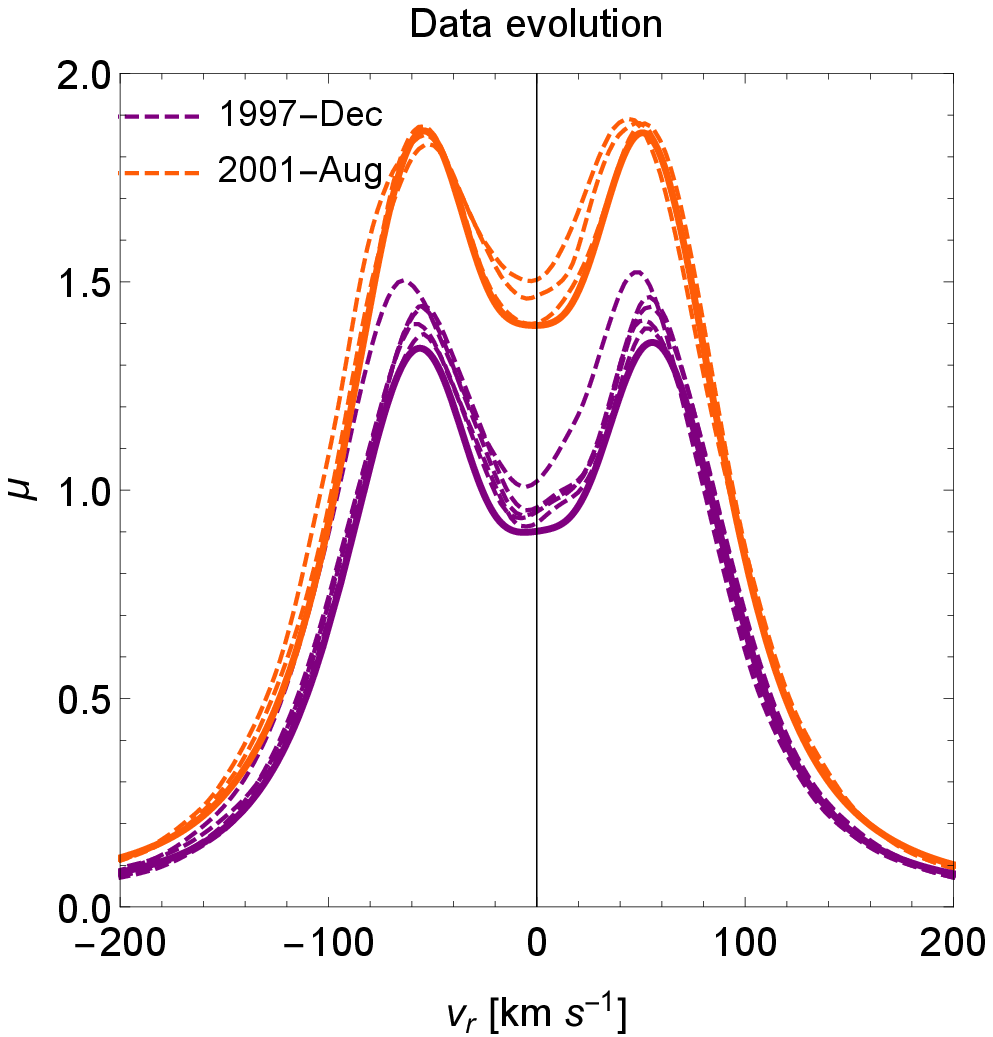}
\includegraphics[width=0.35 \textwidth]{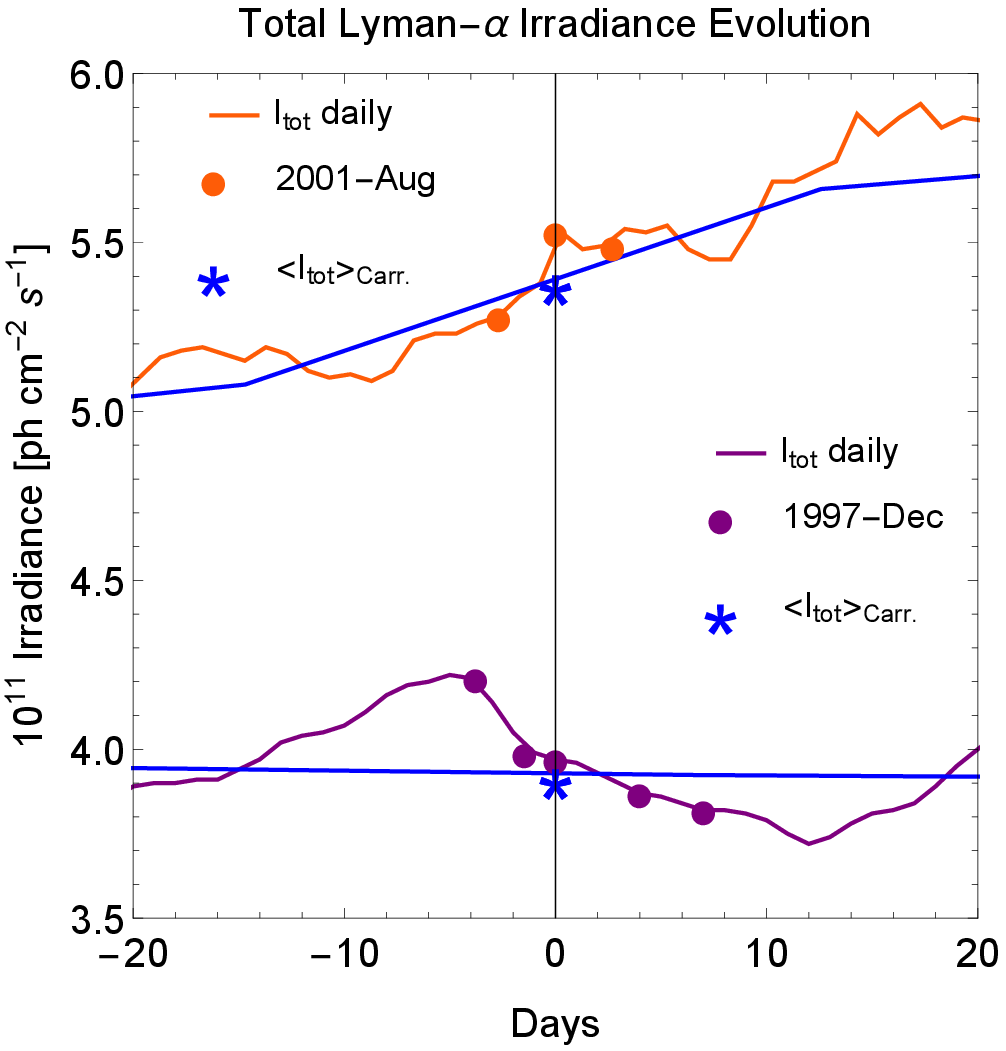}
\caption{
%{\em{fig:obsEvolution}} 
Evolution of the profiles observed within one Carrington rotation period, compared with the model profiles for these Carrington periods (left panel). The dashed orange and purple lines are the observed profiles. The observations were taken during low solar activity, in Dec 1997 (Dec 2nd, Dec 4th, Dec 6th, Dec 10th, Dec 13th, represented by the purple set of lines). The orange group of profiles was observed during high solar activity, in Aug 2001 (Aug 22nd, Aug 24th, Aug 27th). The thick lines in the corresponding colors present the simulated profiles obtained by using the Carrington-averaged value of $I_{\mathrm{tot}}$ interpolated for the date of Dec 6th (lower plot in the left panel) and Aug 27th (upper plot in the left panel). The right panel shows the corresponding daily total flux in Lyman-$\alpha$ $I_{\mathrm{tot}}$ (purple and orange lines), the daily values of $I_{\mathrm{tot}}$ for the observed profiles (purple and orange dots), and the Carrington period-averaged value of the total flux (blue stars), for which the model profiles shown in the left panel were calculated. The blue line represents linear interpolation  between Carrington period averages of the LASP daily $I_{\mathrm{tot}}$ values, which are part of the time series recommended for use with our model (cf. the blue line in Figure~\ref{fig:totalFlux}). The horizontal scale in the right-hand panel is in days prior to or after the center dates for the two sets of lines, i.e., December 6-th 1997 (for the purple set) and Aug. 27-th, 2001, (for the orange set). The time span shown includes $\pm 3/4$ Carrington periods.}
\label{fig:obsEvolution}
\end{figure*}

For modeling radiation pressure for neutral H and D atoms in the heliosphere at an arbitrary longitude and latitude in the heliosphere and an arbitrary time $t$ we recommend to use the Lyman-$\alpha$ total irradiance averaged over Carrington rotations, linearly interpolated to $t$, and not the corresponding daily value of $I_{\mathrm{tot}}$ extracted from the daily time series from LASP. 
The reason for this is the following. For a given date, the value of $I_{\mathrm{tot}}$ from the LASP time series is only valid for the geometric location of the Earth. The instantaneous irradiance for other longitudes is unknown. Using the daily values directly from the LASP time series would imply that the entire Sun is fluctuating in its Lyman-$\alpha$ output precisely as measured by LASP. However, from the fact that the measured daily $I_{\mathrm{tot}}$ values feature quasi-periodic variation during Carrington period (see the right panel in Figure~\ref{fig:obsEvolution}) it follows this is not the case: most of those variations are due to active regions, which are visible from the observer's hemisphere, but are invisible from (and thus cannot affect) the opposite hemisphere. These active regions change in strength and gradually fade out, while other ones appear in the solar disk at various heliolongitudes and heliolatitudes. Therefore, averaging over solar rotation period seems to provide a good balance between the true variations in the solar output and our ignorance of the solar radiation field in the heliosphere away from the Earth. 

The relation of the short-time variations in $I_{\mathrm{tot}}$ to the distribution of active regions, to their locations structured in heliolatitude, one may suspect that radiation pressure will be a function of heliolatitude. This aspect is still poorly investigated, as discussed by \citet{bzowski_etal:13a}. However, observations of the solar corona in EUV lines \citep{auchere:05} and of the heliospheric backscatter glow \citep{pryor_etal:92}, as well as some theoretical considerations \citep{cook_etal:81a}, suggest that $I_{\mathrm{tot}}$ is a mild function of heliolatitude $\phi$. Therefore we follow the recommendation by \citet{bzowski:08a} and adopt the monthly-averaged $I_{\mathrm{tot}}$ value as the following function of heliolatitude $\phi$:
\begin{equation}
\label{eq:helioLatiItot}
I_{\mathrm{tot}}(\phi)=I_{\mathrm{tot}}(0)\sqrt{a_{\mathrm{Lya}} \sin^2 \phi + \cos^2 \phi}
\end{equation} 
with $a_{\mathrm{Lya}} = 0.8$.

\section{Existing phenomenological models of the solar Lyman-$\alpha$ line}
\label{sec:pastModels}

To our knowledge, the first attempt to approximate the shape of the solar Lyman-$\alpha$ line with an analytic formula is due to \citet{fahr:79}, who proposed to express the spectral irradiance as a function of wavelength $\lambda$ by:
\begin{eqnarray}
I_{\mathrm{sol}}(\lambda) &=& I_{\sun}\left[A\exp\!\!\left[-\left(\frac{\lambda - \lambda_0}{\Delta\lambda_A}\right)^2\right]  \right. \\
 & &\left. - B \exp\!\!\left[-\left(\frac{\lambda - \lambda_0}{\Delta\lambda_B}\right)^2\right]\right] ,\nonumber
\label{eq:fahrProfile}
\end{eqnarray}
with the first term approximating the overall Gaussian shape of the profile with the width $\Delta \lambda_A$ and the second term corresponding to the self-reversal, approximated by another Gaussian shape with the width $\Delta \lambda_B < \Delta \lambda_A$. In this approximation the spectral irradiance $I_{\mathrm{sol}}(\lambda)$, and thus radiation pressure, is a linear function of the total irradiance $I_{\sun}$. An identical bi-Gaussian form (with a more elaborate normalization factor) was also used by \citet{scherer_etal:00a}.

\citet{chabrillat_kockarts:97} used a model with three Gaussian functions:
\begin{equation}
I_{\mathrm{sol}} = I_{\sun} \sum \limits_{i = 1}^{3} \frac{A_i}{s_i \sqrt{2\pi}}\exp\!\!\left[ \frac{(\lambda - \lambda_i)^2}{2 s_i^2}\right]
\label{eq:chabrilProfile}
\end{equation}
to fit the observations by \citet{lemaire_etal:78a}, carried out from inside of the Earth's exosphere. Here, the line profile is composed of two Gaussian representations of the two horns, each horn offset from the line center $\lambda_0$ by $\lambda_i - \lambda_0$ and having a width of $s_i$ (note the difference in the normalization constants for the profiles from \citet{fahr:79} on one hand and from \citet{chabrillat_kockarts:97} on the other hand). The third Gaussian component in Equation~\ref{eq:chabrilProfile} corresponds to the exospheric absorption and should be absent from the profile utilized to model the radiation pressure acting on ISN H. 

The TB09 model of the radiation pressure acting on H atoms due to solar Lyman-$\alpha$ emission involved three Gaussian functions:
\begin{eqnarray}
\label{eq:tarnopolskiProfile}
		\mu \left( v_r, I_{\mathrm{tot}} \right) & =& A \left( 1+B I_{\mathrm{tot}} \right) \exp\!\left[ -Cv_r^2\right] \left( 1+ \right. \\
		& &\left. D\exp\!\left[ Fv_r-Gv_r^2\right] +H\exp\!\left[ -Pv_r-Qv_r^2\right] \right)\nonumber, 
\end{eqnarray}
where $v_r$ is radial velocity of the atom with respect to the Sun. In this model, the radiation pressure $\mu$ factor is a linear function of total irradiance in Lyman-$\alpha$ line $I_{\mathrm{tot}}$. 

This model was modified by \citet{katushkina_etal:15b} to increase the horn-to-minimum ratio. These authors introduced an ad hoc parameter $\gamma$ and suggested the following formula:
\begin{eqnarray}
\label{eq:katushkinaProfile}
\mu \left( v_r,t \right) & = & \mu_0 \left(t \right) \frac{\exp\left(-C v_r^2\right)}{1 + (D + H)(1 + \gamma)} \left[1 + (1 + \gamma) \right. \times \\
 & \times & \left. (D \exp[F v_r - G v_r^2]+ H \exp[-P v_r - Q v_r^2]) \right],\nonumber \\
\mu_0 \left(t \right) & = & 0.64 \times 10^{11} \frac{\left(I_{sol}\left(t \right) \times 10^{-11} \right)^{1.21}}{I_{sol,0}},\nonumber \\
\end{eqnarray}
with $I_{sol,0}=3.32 \times 10^{11}$ ph s$^{-1}$ cm$^{-2}$ \AA$^{-1}$, $C = 3.8312\times 10^{-5}$, $D = 0.73979$, $F = 4.0396 \times 10^{-2}$, $G = 3.5135 \times 10^{-4}$, $H = 0.47817$, $P = 4.6841 \times 10^{-2}$, $Q = 3.3373 \times 10^{-4}$.

For $\gamma = 0$ the profile is identical to that from TB09 and for $\gamma > 1.5$ further increase in the horn-to-minimum ratio becomes saturated. 

\section{New model of the spectral shape of the solar Lyman-$\alpha$ line (IKL)} 
\label{sec:modelIKL}

\subsection{Model construction}
In this paper, we propose a more physical approach to the problem of defining an analytic approximation of the line profile. The shape of the line is determined by the velocity distribution of H atoms that are excited to the second energy level. Recent studies \citep{jeffrey_etal:17a} have shown that in the absence of thermodynamical equilibrium the atom velocity distribution can be described by the kappa distribution and the resulting line profiles are also kappa-like. Hence, in our model we used a kappa function to reproduce the main shape of the Lyman-$\alpha$ emission line:
\begin{equation}
\label{eq:kappa}
F_{K}=A_K \left[1+\frac{(v_r-\mu_K)^2}{2 \sigma_K^2 \kappa}\right]^{- \kappa -1}.
\end{equation}
Here, $\mu_K$ represents radial velocity corresponding to the central wavelength of the line, which we allow to differ from the nominal wavelength of 121.567~nm, $\kappa$ is the $\kappa$ parameter of the kappa-like line profile, and $\sigma_K$ represents the overall width of the profile.

The core of the Lyman-$\alpha$ line is affected by the absorption in the solar transition region. To include this effect in our model, we added a Gaussian component
\begin{equation}
\label{eq:gauss}
F_R=\frac{A_R}{\sigma_R\sqrt{2\pi}} \exp{\left[-\frac{(v_r-(\mu_K+d\mu))^2}{2\sigma_R^2}\right]}
\end{equation}
centered at the wavelength corresponding to radial velocity $\mu_K + d\mu$, i.e., we allow the central wavelength of the Lyman-$\alpha$ line and that of the central reversal to vary by $d\mu$. The width of the central reversal is given by $\sigma_R$. 

Finally, we took into account the fact that the solar spectrum is not flat and we included a linear spectral background: 
\begin{equation}
\label{eq:bkg}
F_{bkg}=a_{bkg} v_r +b_{bkg}.
\end{equation}

\begin{figure}[!h]
\centering
\includegraphics[width=\columnwidth]{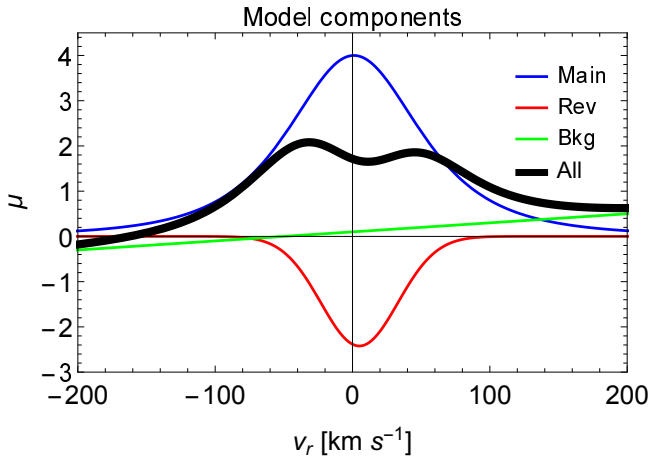}
\caption{
%{\em{fig:funModel}} 
Schematic representation of the components given in Equations~\ref{eq:kappa} through \ref{eq:bkg} of the function given in Equation~\ref{eq:all} used to approximate the observed profiles of the solar Lyman-$\alpha$ line observed by L15. Blue line is the main shape of the line, represented by the kappa function, red line is the central reversal modeled as the Gaussian function, green line is the linear spectral background, and thick black line is the final function. 
%The values of the parameters used in the plot are the following: $A_K = 4$, $\mu_K =1$, $\sigma_K = 65$, $\kappa = 0.8$, $A_R =170$, $d\mu= 4$, $\sigma_R = 28$, $b_{bkg} = 0.1$, $a_{bkg} = 0.002$.
}
\label{fig:funModel}
\end{figure}

The final form of the function is given by 
\begin{equation}
\label{eq:all}
F_{model}=F_K-F_R+F_{bkg}.
\end{equation}
It is schematically presented in Figure~\ref{fig:funModel} along with its components. The relation between the wavelength in the absolute units and the radial velocity is given by
\begin{equation}
\label{eq:lambdaVsvr}
v_r(\lambda)=-\frac{\Delta \lambda}{\lambda_0}c,
\end{equation}
 where $\Delta \lambda = \lambda - \lambda_0$ is the value of the shift from the center of line (in the original observations), $\lambda_0=121.567$~nm is the wavelength of the Lyman-$\alpha$ line, and $c$ is the speed of the light.
Note that negative radial velocities correspond to longer wavelengths. In consequence, the blue horn, that was located on the blue side (left part of Figure \ref{fig:obsLemaire}), is on the right hand side of Figure \ref{fig:funModel}, where we are using radial velocity units.

With the model given by Equations~\ref{eq:kappa} through \ref{eq:all} we were able to reproduce the characteristic shape of the line profile, with two peaks and the asymmetries that are visible in the observed profiles, especially the asymmetry between the heights of the two horns. Further refinement of the model is discussed in the following section.

\subsection{Model fitting}
\label{sec:modelIKL:fitting}
We found that parameter values obtained in fits to individual profiles showed a systematic dependence on $I_{\mathrm{tot}}$, which had to be recognized and accounted for in the final model.
Thus, we constructed the final model in two steps. First, we fitted the parameters of the model defined in Equations~\ref{eq:kappa}--\ref{eq:all} individually to each of the 43 profiles. Then, we analyzed the relations between the fit parameters and $I_{\mathrm{tot}}$ and inscribed them into the model.

\subsubsection{Fitting individual profiles}
\label{sec:oneProfileFit}
We performed the fitting after converting the original wavelength scale given in nanometers to the Doppler scale in km~s$^{-1}$ using Equation~\ref{eq:lambdaVsvr}, and the spectral irradiance originally given cm$^{-2}$~s$^{-1}$~nm$^{-1}$ to the dimensionless $\mu$ parameter, which represents the fraction of solar gravity compensated by radiation pressure, by dividing the spectral irradiance by the scaling factor sclFctH$=3.34 \times 10^{12}$. The $\mu$ units are very convenient from the viewpoint of calculating the trajectories of H atoms in the heliosphere. For all profiles, we used for fitting the portions of the data corresponding to $\pm 200$~km~s$^{-1}$ around 0. This was done to avoid biasing the results by the far wings of the profiles.

\begin{figure}
\centering
\includegraphics[width=\columnwidth]{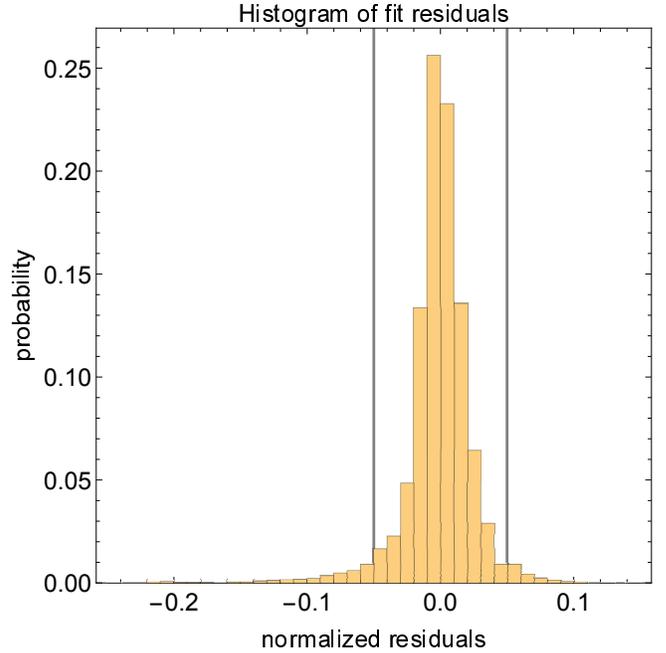}
\caption{
%{\em{fig:resHist}} 
Histogram of the residuals (data - model)/data of model fits to all individual profiles for the radial velocity range $\pm 200$~km~s$^{-1}$. The vertical bars mark the $\pm 5$\% region.}
\label{fig:resHist}
\end{figure}

\begin{figure}
\centering
\includegraphics[width=\columnwidth]{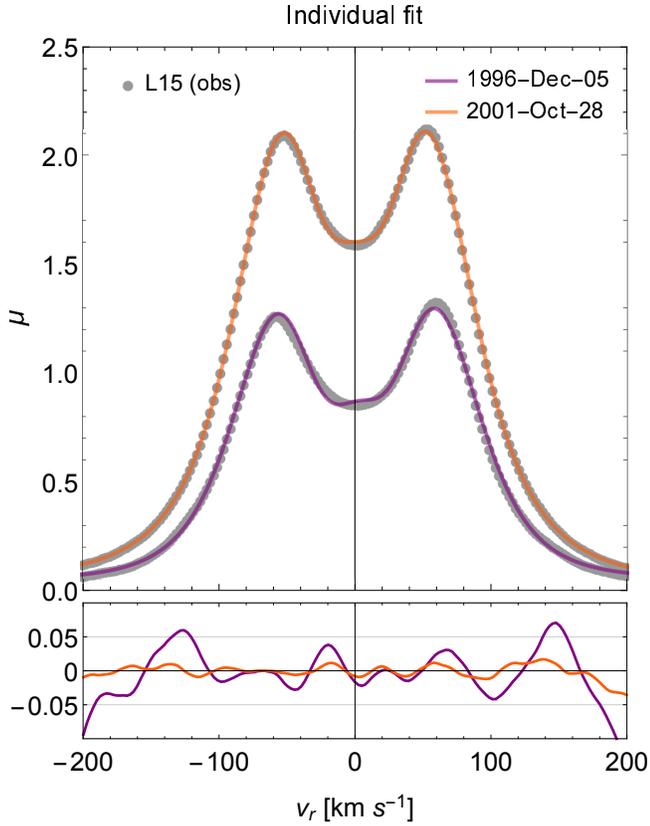}
\caption{
%{\em{fig:fitExmp}} 
Result of the fit of the model defined in Equations~\ref{eq:kappa}--\ref{eq:all} to selected individual profiles. Gray points represent the observations from L15, orange line marks an example fitted profile that was observed near solar maximum (Oct 28th, 2001), and purple line draws an example profile taken near solar minimum (Dec 5th, 1996). The bottom panel presents residuals (difference between data and the model divided by the data) for those two profiles.}
\label{fig:fitExmp}
\end{figure}

Each profile was fitted with the 9-parameter function described by Equations from \ref{eq:kappa} to \ref{eq:bkg} using nonlinear least-squares method from Wolfram Mathematica \citep{Mathematica:10.4}. We obtained very good match between the data and the fitted functions in most of the cases. Within $\pm 200$~km~s$^{-1}$, only 5 out of 43 profiles have residuals (defined here as (data - model)/data) larger than 10\%, and this happens only for $v_r$ close to $\pm 200$~km~s$^{-1}$. Generally, most of the residuals are less than 5\%, as shown in Figure~\ref{fig:resHist}. Example profiles for the minimum and maximum of solar activity and the corresponding residuals are shown in Figure \ref{fig:fitExmp}. Near the maximum of solar activity (Oct 28th, 2001) the magnitude of the profile was much higher than during solar minimum (Dec 5th, 1996) and consequently the magnitude of radiation pressure was also larger. The residuals in both cases are safely within the $\pm 5$\% margin for radial velocities $\pm 100$~km~s$^{-1}$, i.e., within an interval the most important for ISN H in the heliosphere. These values are within the data uncertainty level given by L15. 

As a result of fitting individual profiles, we obtained 43 sets of parameters, each describing a different profile observed by the SUMER detector. Their magnitudes and uncertainties are shown in Figure~\ref{fig:linCorr} as a function of $I_{\mathrm{tot}}$. 

\subsubsection{Parameter correlation with $I_{\mathrm{tot}}$}
\label{sec:linCorr}
\begin{figure*}
\centering
\includegraphics[width=\textwidth]{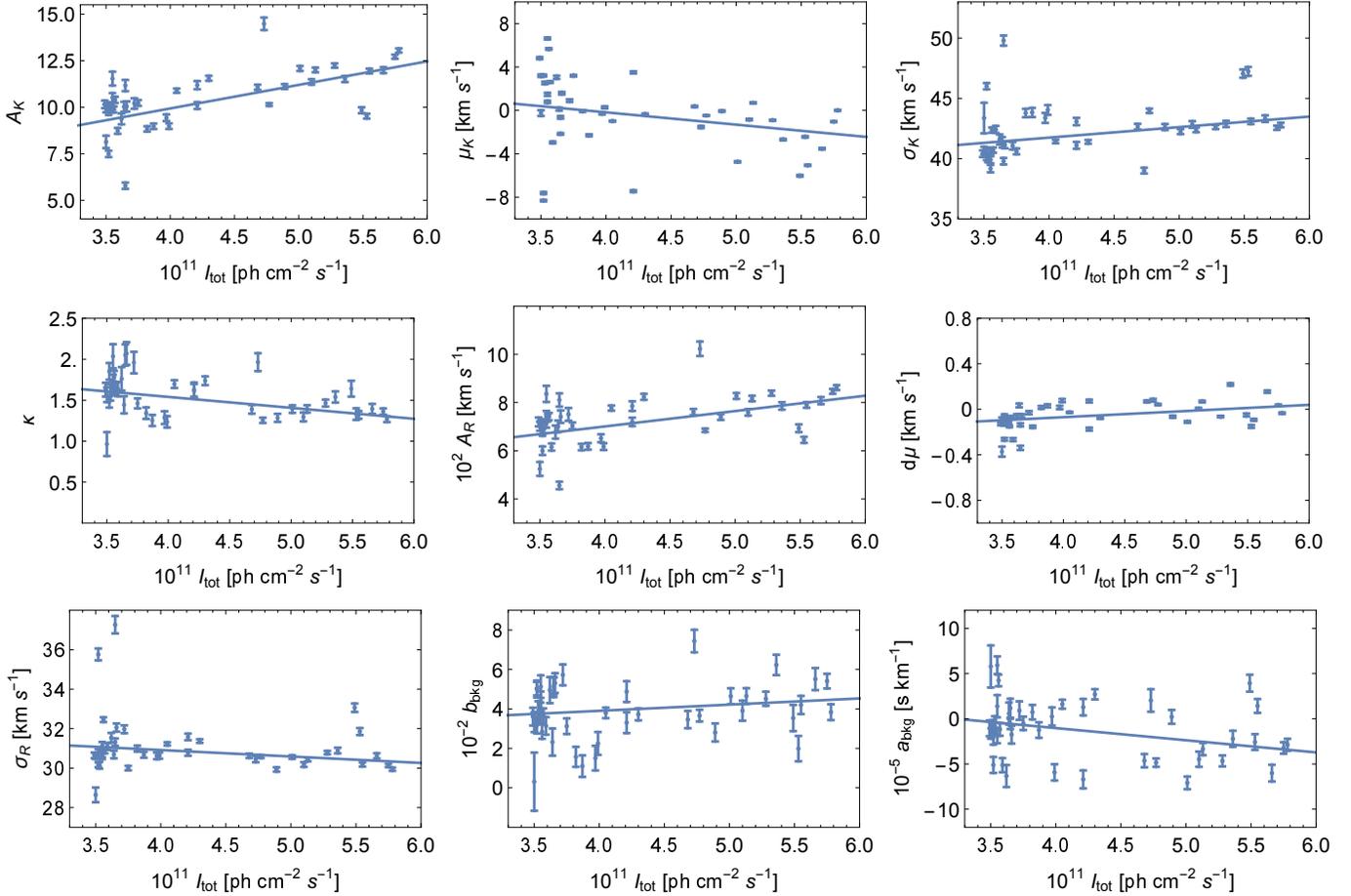}
\caption{
%{\em{fig:linCorr}} 
Correlations between the parameters of the model, defined in Equations~\ref{eq:kappa}--\ref{eq:all} and obtained from fitting individual profiles, and the composite total irradiance in Lyman-$\alpha$. Blue points with error bars (in some cases very short) show the values of best fit parameters to individual profiles. Solid lines are the result of fitting the linear correlations to those points, taking into account errors from the fitting to the observations. Data points with the corresponding errors are available as the Data behind the Figure.}
\label{fig:linCorr}
\end{figure*}

The model parameter values, shown in Figure~\ref{fig:linCorr}, clearly show correlation with the line integrated irradiance $I_{\mathrm{tot}}$, albeit with a large scatter. Since the procedure being used to compute the density and higher moments of the distribution function of ISN H and D in the WTPM model \citep{tarnopolski_bzowski:08a, tarnopolski_bzowski:09} must calculate thousands of trajectories of individual atoms, and each trajectory is calculated by numerically solving Equation~\ref{eq:simpleEqMotion} with the $\mu$ factor being a function of $v_r$ and $t$, our goal was to create a model for $\mu(v_r, I_{\mathrm{tot}}(t))$ that can be expressed by a relatively simple analytical formula. On the other hand, the model must be able to reproduce the true radiation pressure relatively accurately because the density and bulk velocity of ISN H and D are sensitive functions of radiation pressure. 

TB09 used a simple proportionality scaling of the line profile by $I_{\mathrm{tot}}$, which was demonstrated by \citet{schwadron_etal:13a} and \citet{katushkina_etal:15b} to likely be too simple. With more than 40 sets of parameters for profiles observed during the entire solar cycle we could search for relations between profile parameters corresponding to various levels of solar activity. We correlated every parameter of the model ($A_K$, $\mu_K$, $\sigma_K$, $\kappa$, $A_R$, $d\mu$, $\sigma_R$, $a_{bkg}$, and $b_{bkg}$) with the total irradiance in Lyman-$\alpha$ scaled by the solar cycle average value (computed over 23rd solar cycle for which observed profiles are available)  $I_{\mathrm{tot}}/\langle I_{\mathrm{tot}}\rangle$. We found that the correlation of all of the parameters with $I_{\mathrm{tot}}$ can be approximated by a linear function, as shown in Figure~\ref{fig:linCorr}. 

We used this finding in the construction of the final model of $\mu(v_r, I_{\mathrm{tot}}(t))$. We fitted each of the 9 parameters of the model with the linear function using the least-squares method. A parameter $P_i$ is expressed in following convenient form:
\begin{equation}
\label{eq:paramForm}
P_i=\beta_i \left( 1+ \alpha_i \frac{I_{\mathrm{tot}}}{\langle I_{\mathrm{tot}}\rangle} \right),
\end{equation}
where $P_i=(A_K,\mu_K,\sigma_K,\kappa,A_R,d\mu,\sigma,b_{bkg},a_{bkg})$, $\langle I_{\mathrm{tot}}\rangle=4.25 \times 10^{11}$~ph~cm$^{-2}$~s$^{-1}$, and $\alpha_i$ and $\beta_i$ are the fit parameters, listed in Table~\ref{tab:linCo}. In each of those nine fits, the data were the 43 values of a given parameters for the 43 Lyman-$\alpha$ profiles, and the independent variable was the corresponding daily $I_{\mathrm{tot}}/\langle I_{\mathrm{tot}}\rangle$ value. The data were weighted by inverse squares of the parameter errors, obtained from the fits presented in Section~\ref{sec:oneProfileFit}. Results of fitting the parameters from Equation~\ref{eq:paramForm} are listed in Table~\ref{tab:linCo} and the fitted lines shown in Figure~\ref{fig:linCorr}. 

Instead of the composite total irradiance in Lyman-$\alpha$ potentially one could use the solar radio flux F10.7. These two quantities are strongly correlated and the first one is calculated from the latter when there are no direct observations \citep{woods_etal:96a}. We  performed the whole model fitting procedure using radio the flux F10.7 instead of $I_{\mathrm{tot}}$, but the results were a little less accurate than in the case of total irradiance in Lyman-$\alpha$.

\section{Results}
\label{sec:results}
\subsection{Model for Hydrogen}
\label{sec:results:H}
The final result for radiation pressure model for neutral H in the heliosphere is given by Equations~\ref{eq:kappa}--\ref{eq:all} and \ref{eq:paramForm}, summarized in Equation~\ref{eq:modelFinal}, with the coefficients listed in Table~\ref{tab:linCo}. The free parameters are the total irradiance in the Lyman-$\alpha$ line $I_{\mathrm{tot}}$ in ph~cm$^{-2}$~s$^{-1}$ and radial velocity $v_r$ in km~s$^{-1}$. The result is in the units of compensation of the solar gravity force (the $\mu$-units).

Summarizing, the model of radiation pressure $\mu(r, v_r, t, \phi)$ for radial velocity $v_r$ (in km~s$^{-1}$), time $t$, heliolatitude $\phi$, heliocentric distance $r$ is given by the following set of equations
\begin{eqnarray}
\label{eq:modelFinal}
I_{\mathrm{tot}}(t, 0) & = & I_{\mathrm{tot},n} + \frac{I_{\mathrm{tot},n+1} - I_{\mathrm{tot},n}}{t_{n+1} - t_n}(t - t_n) \nonumber \\
I_{t,\phi} & = & I_{\mathrm{tot}}(t, 0)\sqrt{a_{\mathrm{Lya}} \sin^2 \phi + \cos^2 \phi} \nonumber \\
P_i & = & \beta_i\left(1 + \alpha_i \frac{I_{t,\phi}}{\langle I_{\mathrm{tot}}\rangle}\right) \nonumber \\
v_r & = & -\frac{\lambda - \lambda_0}{\lambda_0} c \\ 
F_K & = & A_K \left[1+\frac{(v_r-\mu_K)^2}{2 \sigma_K^2 \kappa}\right]^{- \kappa -1} \nonumber \\
F_R & = &\frac{A_R}{\sigma_R\sqrt{2\pi}} \exp{\left[-\frac{(v_r-(\mu_K+d\mu))^2}{2\sigma_R^2}\right] } \nonumber \\
F_{bkg} & = & a_{bkg} v_r +b_{bkg} \nonumber \\
\mu(r, v_r, t, \phi) & = & (F_K -F_R + F_{bkg}) (r_E/r)^2 \nonumber,
\end{eqnarray}
where $t_n, t_{n+1}$ are the times of the starts of the Carrington rotations immediately preceding and following the time of calculation $t$, $a_{\mathrm{Lya}}=0.8$, $I_{\mathrm{tot},n}$; $I_{\mathrm{tot},n+1}$ are the values of the LASP time series averaged over the $n$-th and $n+1$-th Carrington periods, respectively; $\langle I_{\mathrm{tot}}\rangle = 4.25 \times 10^{11}$~ph~cm$^{-2}$~s$^{-1}$; and $P_i = (A_K,\mu_K,\sigma_K,\kappa,A_R,d\mu,\sigma,b_{bkg},a_{bkg})$ are the parameters of the profile, obtained from Table~\ref{tab:linCo} for H and Table~\ref{tab:linCoD} for D. 

\begin{deluxetable}{ccc}
\tablecaption{\label{tab:linCo}Coefficients of the linear correlations between the model parameters for H and total irradiance in Lyman-$\alpha$, defined in Equation~\ref{eq:paramForm}.}
\tablehead{
		\colhead{Parameter ($P_i$)} & \colhead{$\beta_i$} & \colhead{$\alpha_i$}
	}
\startdata
$A_K$ & $  4.873$ & $1.10341$ \\
$\mu_K$ & $  4.349$ & $-1.1064$ \\
$\sigma_K$ & $ 38.251$ & $0.0969713$ \\
$\kappa$ & $ 2.074$ & $-0.273553$ \\
$A_R$ & $ 446.14$ & $0.606729$ \\
$d\mu$ & $ -0.285$ & $-0.804873$ \\
$\sigma_R$ & $  32.224$ & $-0.0431647$ \\
$b_{bkg}$ & $ 0.026$ & $0.503516$ \\
$a_{bkg}$ & $0.435\cdot 10^{-4}$ & $-1.3148$ \\
\enddata
\tablecomments{ Model along with all parameters is available online: http://users.cbk.waw.pl/~ikowalska/index.php?content=lya}
\end{deluxetable}

\subsection{Model for Deuterium}
\label{sec:results:D}
\begin{figure}
\centering
\includegraphics[width=1.0\columnwidth]{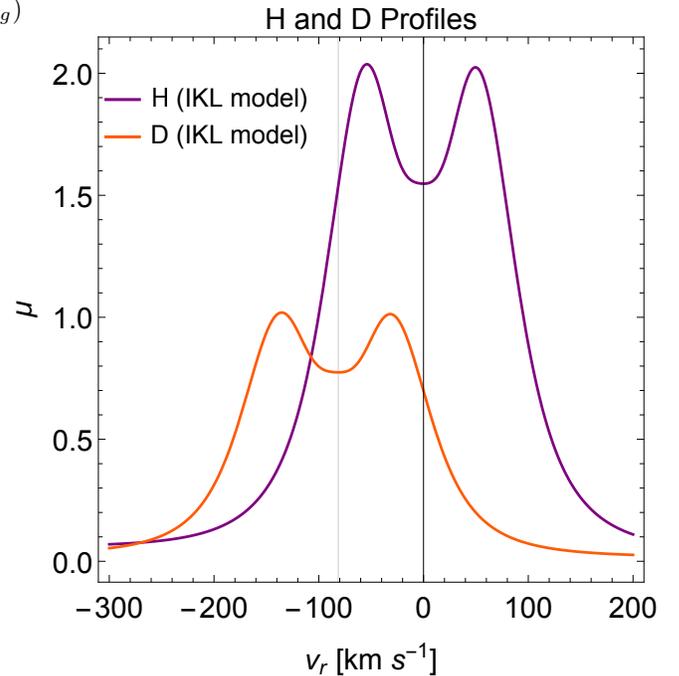}
\caption{
%{\em{fig:profileHD}} 
Comparison between the profile (Oct 28th, 2001) obtained by using our new model for Hydrogen (purple line) and Deuterium (orange line). The D profile is shifted by $v_r \rightarrow v_r + 81.3802$~km~s$^{-1}$ due to isotope effect and scaled by the factor of 0.5003 due to the mass difference between D and H.}
\label{fig:profileHD}
\end{figure}

The same formulae can be used to model radiation pressure for neutral D. This is a result of a simple shift in the wavelength due to isotope effect ($v_r \rightarrow v_r + 81.3802$~km~s$^{-1}$) and scaling down by the H/D mass ratio, equal to 0.5003. The coefficients thus obtained are given in Table~\ref{tab:linCoD} and example profile (Oct 28th, 2001) presented in Figure \ref{fig:profileHD} along with a H profile for comparison.

\begin{deluxetable}{ccc}
\tablecaption{\label{tab:linCoD}Coefficients of the linear correlations between the model parameters for D and total irradiance in Lyman-$\alpha$, defined in Equation~\ref{eq:paramForm}.}
\tablehead{
		\colhead{Parameter ($P_i$)} & \colhead{$\beta_i$} & \colhead{$\alpha_i$}
	}
\startdata
$A_K$ & $  2.438$ & $1.10341$ \\
$\mu_K$ & $-77.031$ & $0.196728$ \\
$\sigma_K$ & $ 38.251$ & $0.0969713$ \\
$\kappa$ & $ 2.074$ & $-0.273553$ \\
$A_R$ & $ 223.24$ & $0.606729$ \\
$d\mu$ & $ -0.285$ & $-0.804873$ \\
$\sigma_R$ & $  32.224$ & $-0.0431647$ \\
$b_{bkg}$ & $ 0.013$ & $0.503516$ \\
$a_{bkg}$ & $0.218\cdot 10^{-4}$ & $-1.3148$ \\
\enddata
\tablecomments{Model along with all parameters is available online: http://users.cbk.waw.pl/~ikowalska/index.php?content=lya}
\end{deluxetable}

\section{Discussion}
\label{sec:discussion}

\subsection{Variation of the Lyman-$\alpha$ profile during solar rotation}
\label{sec:discussion:solRot}

The data from L15 offer an opportunity to study the variation of the full-disk profiles within a solar rotation. We have identified two time intervals when at least three profiles are available within a time interval less than the Carrington rotation period: one during low solar activity (in December 1997, with 5 profiles), and the other one during high solar activity (in August 2001, 3 profiles). These profiles are shown in Figure~\ref{fig:obsEvolution} along with our model profile, calculated for $I_{\mathrm{tot}}$ equal to the mean value over the 27.4 day interval centered at the arithmetic mean of the dates from the data corresponding profile subsets. 

The full-disk profiles vary considerably practically overnight, both during high and, surprisingly, low solar activity. These variations are reflected in the total irradiance $I_{\mathrm{tot}}$. This finding illustrates why it is not realistic to expect that a profile measured for any individual date will be representative for the entire heliolongitude range. It also illustrates that adopting a model where the value of $I_{\mathrm{tot}}$, which is the driver for the model, obtained from interpolation of Carrington-period averages of the daily LASP series is a reasonable choice. The scatter between the profiles observed within a few days from each other may be adopted as a measure of the uncertainty of our model resulting from adopting average profiles characteristic for a the entire range of heliolongitudes.  

The quality of the approximation we propose is further illustrated in Figure~\ref{fig:profReconstruction}. We show that since in our approach the sole model parameter is the disk-integrated solar irradiance $I_{\mathrm{tot}}$, it is reasonable to expect that average profiles, obtained from measurements with similar values of $I_{\mathrm{tot}}$, will be well reproduced by the model evaluated with $I_{\mathrm{tot}}$ equal to the mean value of this parameter for all profiles within such a group. To reduce statistical scatter we decided to split the observed data set into three classes according to $I_{\mathrm{tot}}$ values: low activity ($I_{\mathrm{tot}} \leq 4 \times 10^{11}$~cm$^{-2}$~s$^{-1}$), medium activity ($4 < \times 10^{11} < I_{\mathrm{tot}} \leq 5\times 10^{11}$~cm$^{-2}$~s$^{-1}$), and high activity ($I_{\mathrm{tot}} > 5\times 10^{11}$~cm$^{-2}$~s$^{-1}$), and compare the model profiles with the observed profiles averaged over the three subsets of $I_{\mathrm{tot}}$. We also show profiles that are obtained for the aforementioned three subsets from averaging the individually fit profiles, discussed in Section~\ref{sec:oneProfileFit}. The average individually fit profiles agree very well with the average observed profiles. The agreement between the averaged observed profiles and the profiles obtained from the final model is also good, but a little worse than for individually fit profiles. This is because in the final model we use the linear dependence of the fit parameters on the total irradiance. 

\subsection{Horn to minimum ratios}
\label{sec:horn2min}
One of important aspects of the Lyman-$\alpha$ line profile is the depth of the central reversal and the height of the two horns. It can be represented as ratios of the spectral irradiance for one of the horns to the local minimum in the spectral irradiance, located between the two horns. \citet{schwadron_etal:13a} and \citet{katushkina_etal:15b} suggested that the ratio of the blue horn to the reversal depth needs to be larger than that in the model by TB09 to reproduce the spectrum of interstellar neutral H flux observed by IBEX-Lo. 

\begin{figure*}
\centering
\includegraphics[width=0.45 \textwidth]{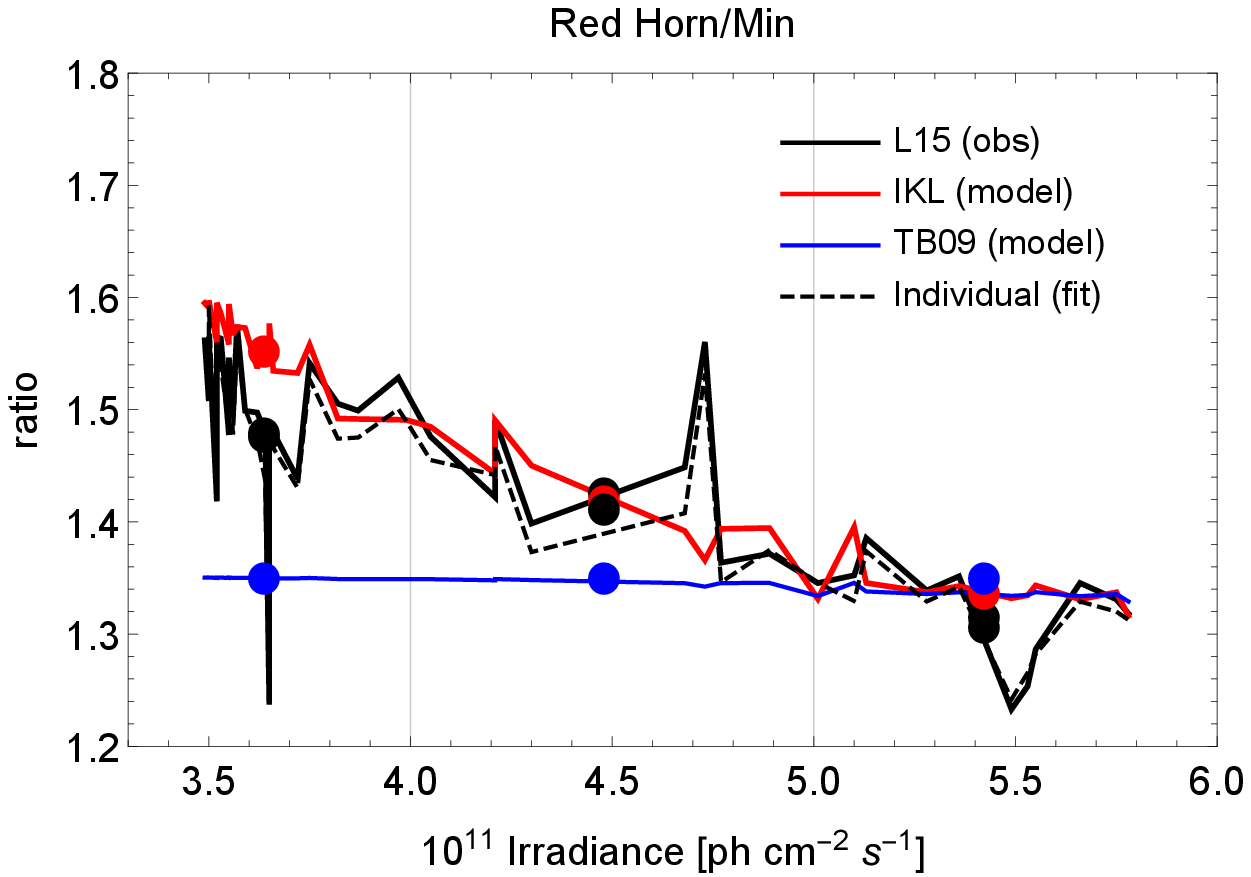} 
\includegraphics[width=0.45 \textwidth]{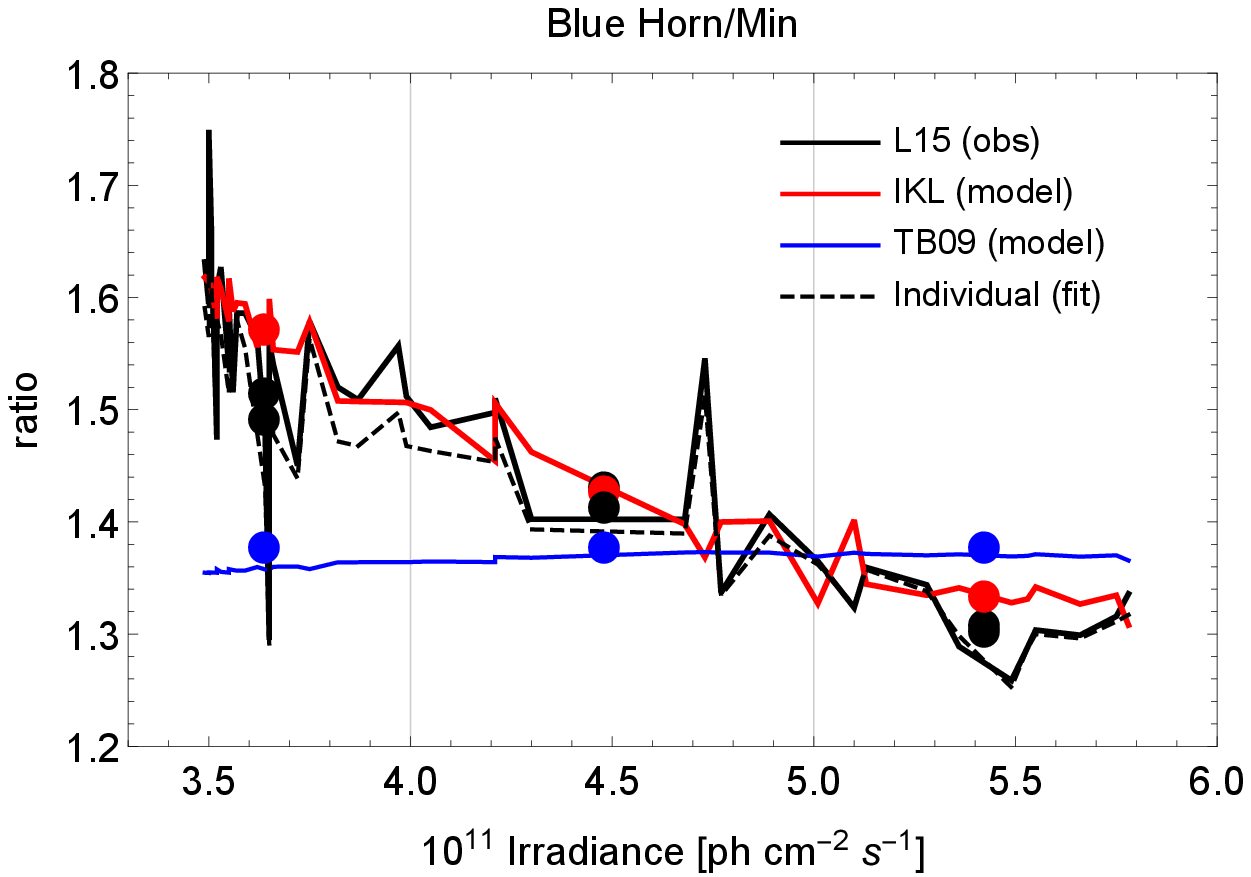}
\caption{
%{\em{fig:ratioBCRC}} 
Ratios of spectral irradiances for the red horn to the central reversal (left panel) and the blue horn to central reversal (right panel) from the observations by L15 (black lines) and from our fits to individual profiles (dashed black), shown as a function of the total irradiance $I_{\mathrm{tot}}$. The ratios obtained from the final model are drawn with the red lines, and the corresponding ratios from the model by TB09 are illustrated with the blue lines. The three sets of color dots mark the respective horn to minimum ratios for the data and the model profiles averaged over $I_{\mathrm{tot}} < 4.0\times 10^{11}$, $4.0\times 10^{11} \le I_{\mathrm{tot}} \le 5.0\times 10^{11}$, and $I_{\mathrm{tot}} > 5.0\times 10^{11}$. }
\label{fig:ratioBCRC}
\end{figure*}

The horn-to-minimum ratios for all observed profiles are shown in Figure~\ref{fig:ratioBCRC}, separately for the blue (positive velocities) and red (negative velocities) horns, as a function of the total irradiance $I_{\mathrm{tot}}$. Clearly, this ratio drops with the increase of solar activity approximately linearly, i.e., the depth of the central reversal approximately linearly decreases with the increase of solar activity. This can be understood since on one hand, \citet{tian_etal:09c} showed that the Lyman-$\alpha$ profiles observed in active regions do not feature considerable self-reversal, and on the other hand, the number of active regions visible on the solar disk increases with the increase of solar activity, so their contribution to the disk-averaged profiles should be larger than during low solar activity. 

The present model does show this dependence of the horn-to-minimum ratios on the total irradiance. The magnitude of the model horn-to-minimum ratios are in agreement with those observed, for both the red and blue horns, and as well for the models fit to individual profiles as for the global model presented in Section~\ref{fig:linCorr}, where the model parameters are linear functions of $I_{\mathrm{tot}}$. This also holds for the model and data profiles averaged within the three aforementioned data subsets. 

\subsection{Total irradiance vs radiation pressure at line center}
\label{sec:muAtLineCtr}
\begin{figure}
\centering
\includegraphics[width=0.45 \textwidth]{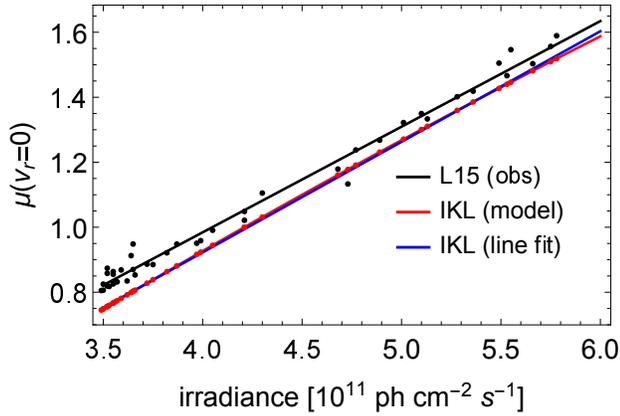}
\caption{
%{\em{fig:CentralMu}}
Radiation pressure at the line center (i.e., for H atoms with radial velocity equal to 0) as a function of the total irradiance $I_{\mathrm{tot}}$, obtained from the profiles observed by L15 (black dots) and a linear function fitted to these data (black line). Red dots mark the center of line radiation pressure calculated from the model for total irradiance values corresponding to those from the data. The red line is the radiation pressure at the line center obtained directly from our model, and the blue line is a linear function fit to this model.}
\label{fig:CentralMu}
\end{figure}
Another aspect is the relation between $I_{\mathrm{tot}}$ and the level of radiation pressure at the line center: $\mu_0\left(I_{\mathrm{tot}} \right) = \mu\left(v_r = 0, I_{\mathrm{tot}} \right)/I_{\mathrm{tot}}$. Historically, this is an important parameter because it was used to assess the radiation pressure from the wavelength-integrated intensity in the models where the dependence of radiation pressure on $v_r$ was neglected. 

In the past, this ratio was modeled using a nonlinear relation. \citet{vidal-madjar_phissamay:80} suggested that the ratio between the spectral flux and line-integrated intensity is $f_0 = 0.54 \times I_{\mathrm{tot}}^{1.54}$. \citet{emerich_etal:05} modified this relation to $f_0 = 0.64 \times I_{\mathrm{tot}}^{1.21}$, and \citet{lemaire_etal:05} to $f_0 = 0.62 \times I_{\mathrm{tot}}^{1.23}$. In all of these formulae, $I_{\mathrm{tot}}$ was to be used in the units $10^{11}$~cm$^{-2}$~s$^{-1}$. While clearly nonlinear in general, for the values of $I_{\mathrm{tot}}$ characteristic for the Sun these relations are linear in a very good approximation, which was pointed out by \citet{bzowski_etal:13a} and \citet{lemaire_etal:15a}.

In our model, the quantity $\mu_0(I_{\mathrm{tot}})$ can be calculated analytically:
\begin{equation}
\label{eq:muVr0}
\mu_0 = b_{bkg} + A_K \left(1 + \frac{\mu_K^2}{2 \kappa \sigma_K^2} \right)^{-1 - \kappa} - \frac{A_R}{\sqrt{2 \pi}\sigma_R}\exp\!\left[-\frac{(\mu_K + d\mu)^2}{2 \sigma_R} \right],
\end{equation}
where the coefficients must be calculated for a given value of $I_{\mathrm{tot}}$ from Equation~\eqref{eq:paramForm} using the appropriate parameter values from Table~\ref{tab:linCo}. A plot of this formula is shown in Figure~\ref{fig:CentralMu}, along with the $\mu_0$ ratios taken directly from the data and with a linear fit to Equation~\eqref{eq:muVr0}. Clearly, the linear approximation for the dependence of the radiation pressure at the line center on the total irradiance is a very good one, both for the data and for our model. Our model systematically underestimates the radiation pressure in comparison with the observed profiles by $\sim 9$\% for the lowest values of $I_{\mathrm{tot}}$ and by $\sim 3$\% for $I_{\mathrm{tot}} > 5.2 \times 10^{11}$~cm$^{-2}$~s$^{-1}$. Therefore, for the models neglecting the dependence of radiation pressure on radial speed of H atoms we recommend using the appropriately scaled formula from L15. We point out, however, that ISN H atoms in the heliosphere spend very little time at $v_r \simeq 0$, so this systematic difference is not expected to significantly affect heliospheric models. Note that this systematic difference vanishes for larger radial speed, as can be inferred from the accuracy of the reproduction of the the horn/minimum ratio (see Figure~\ref{fig:ratioBCRC}).

\subsection{Comparison with TB09}
\label{sec:compareTB09}
\begin{figure}
\centering
\includegraphics[width=0.8\columnwidth]{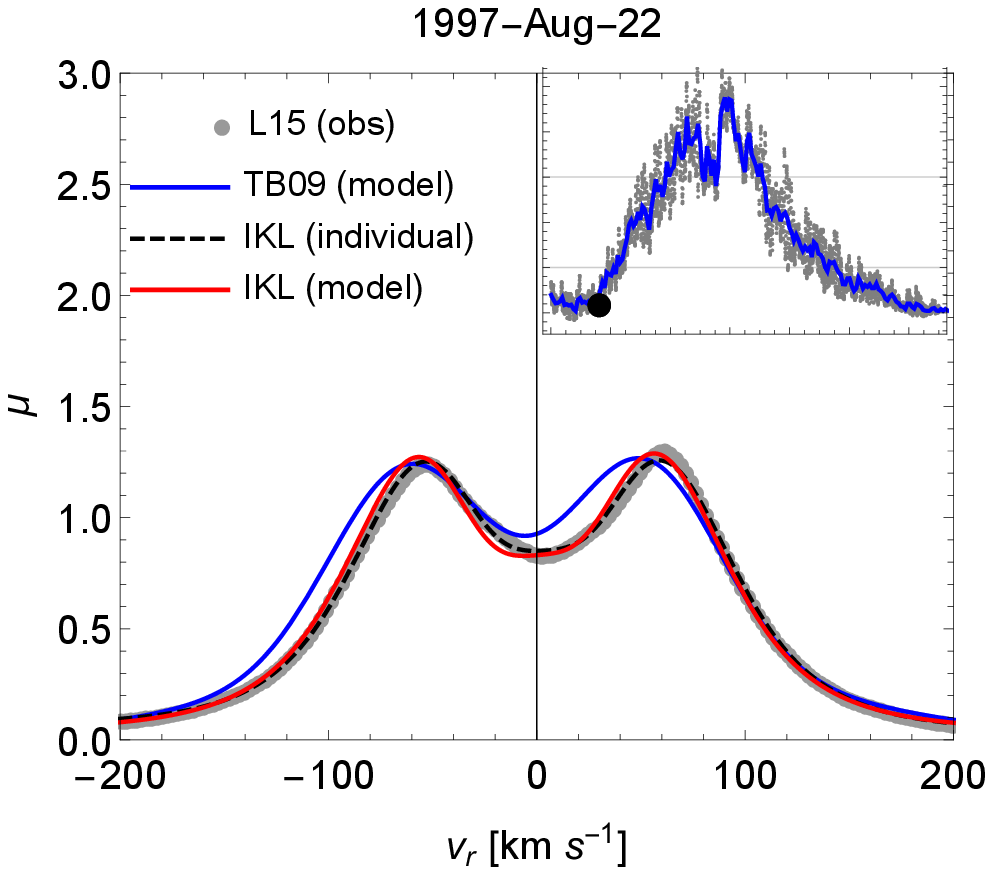}

\includegraphics[width=0.8\columnwidth]{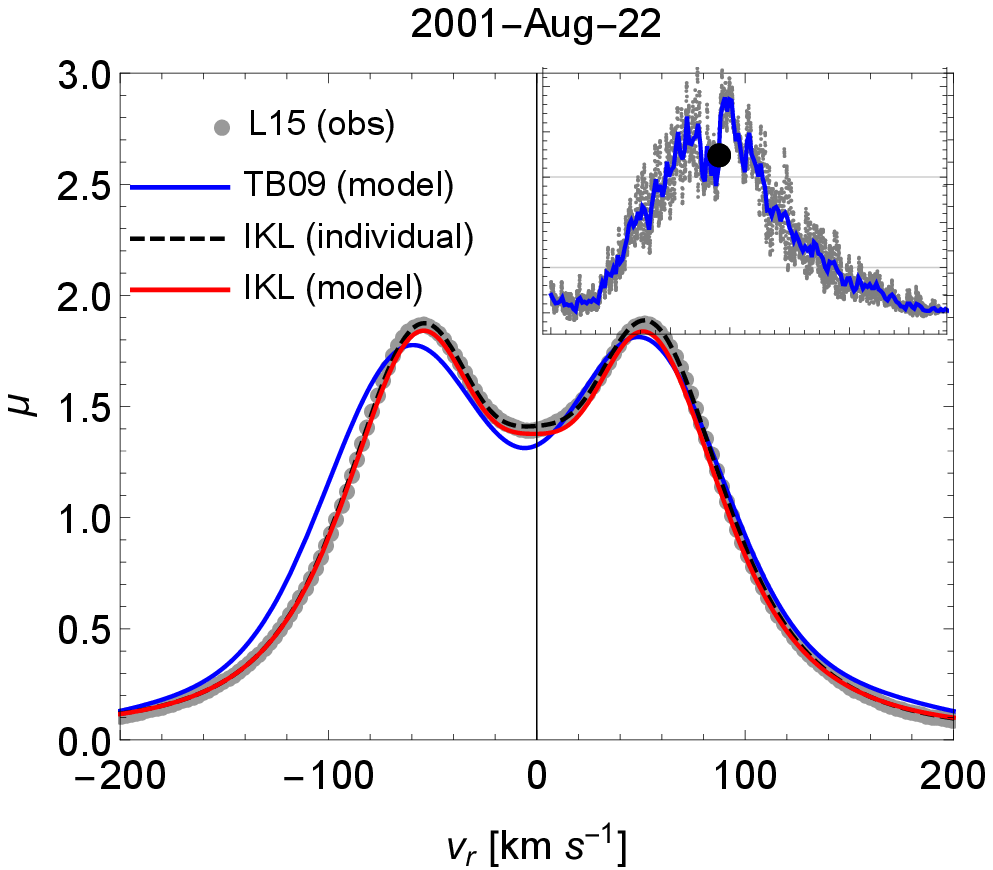}

\includegraphics[width=0.8\columnwidth]{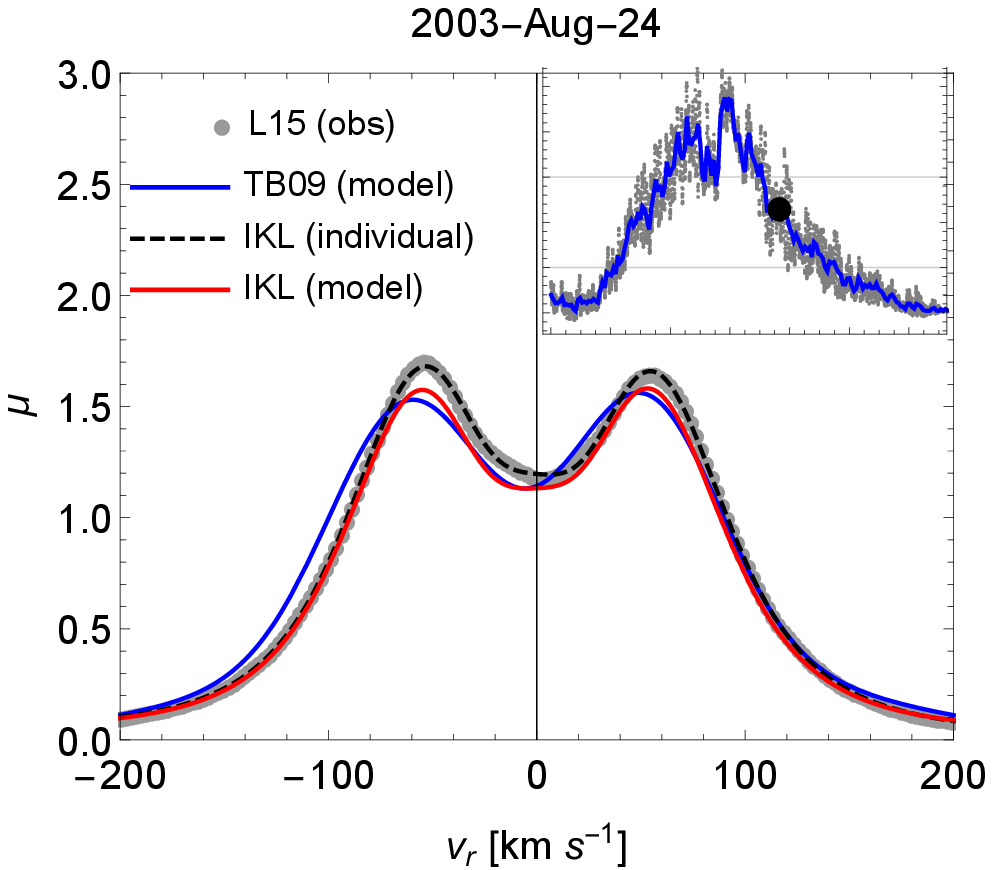}

\caption{
%{\em{fig:profReconstruction}} 
Comparison of observed profiles with the model for example dates (shown in panel headings) for low (upper panel), high (middle panel), and intermediate (lower panel) level of solar activity. Gray dots represent the observed profiles. The broken lines represent fits to the individual profiles (Section~\ref{sec:oneProfileFit}), the red lines represent the final model calculated for the recommended values of the total solar irradiance $I_{\mathrm{tot}}$, i.e., Carrington averages linearly interpolated to the times of observations (Section~\ref{sec:results:H}). The blue lines represent the TB09 model, evaluated for the same values of $I_{\mathrm{tot}}$ as the red lines. The insets are similar to Figure~\ref{fig:totalFlux} with the selected date of observation marked by thick dots. The horizontal lines in the insets mark the division of $I_{\mathrm{tot}}$ values into the three categories of low, intermediate, and high solar activity.}
\label{fig:profReconstruction}
\end{figure}

The present model differs from TB09 in some important aspects. The model function that we used is much more sophisticated than the simple three-Gaussian model used by TB09 and the quality of data reproduction that we have obtained is generally better. This is understandable, since the present model was obtained from many more profiles than available to TB09. The differences between the two models are illustrated in Figure~\ref{fig:profReconstruction}. 

Due to the definition of the model function, TB09 has the horn-to-minimum ratios almost independent of the total irradiance $I_{\mathrm{tot}}$. These ratios are illustrated in Figure~\ref{fig:ratioBCRC}. Clearly, TB09 underestimates the horn-to-minimum ratios for the solar minimum conditions, as suggested by \citet{schwadron_etal:13a} and \citet{katushkina_etal:15b}. It is not clear at the moment if this profile change will be sufficient to understand the spectrum of ISN H observed by IBEX, but definitely, the model we have derived agrees in this aspect with the suggestions in those papers.

Effects of the new model of radiation pressure as a function of atom radial velocities are likely the most important for ISN H and D distributions, especially for the low solar activity conditions. The radiation pressure dependence on radial velocity is stronger than in TB09. This will likely affect estimates of the PUIs and solar wind energetic neutral atoms (ENA) production, as well as the heliospheric back-scatter Lyman-$\alpha$ glow. On the other hand, effects for ENA survival probabilities will likely be small since ENAs due to their large velocity are mostly sensitive to the regions of the profile outside the self-reversal. 

The effect for the distribution of ISN D in the inner heliosphere needs careful evaluation, but we believe it will likely be smaller than for ISN H, because ISN D is mostly sensitive to the blue horn region of the Lyman-$\alpha$ profile, and in this region, as can be seen in Figure~\ref{fig:profReconstruction}, the magnitudes of the $\mu$ factor returned by the present model and that of TB09 are similar. 

This discussion certainly must be regarded as speculative. Effects of the new model of radiation pressure on the models of heliospheric ISN H and D and some of their derivative populations (PUIs, solar wind ENAs, direct samples collected by IBEX-Lo) and the helioglow will be a subject of a future paper.

\subsection{Uncertainties of the model and effects of hypothetical bias in the absolute calibration of the solar Lyman-$\alpha$ irradiance}

The uncertainties of our model are a result of superposition of many different errors and uncertainties. They include the accuracy of the absolute calibration of the total irradiance $I_{\mathrm{tot}}$, the accuracy of the measurements by the SUMER detector and of the deconvolution process used by L15, as well as fit errors obtained in our two fitting procedures. We have no knowledge of some of those uncertainties, therefore it is almost impossible to estimate the accuracy of our model in a formal way. 

L15 suggest the total uncertainties of their profiles are at a level $\sim 15$\%. The residuals in our individual profile fits are much less than that (see Figure~\ref{fig:resHist}). However, profiles obtained for specific days are not fully representative for the purpose of modeling the solar radiation pressure for the needs of heliospheric studies. 

The uncertainties related to this aspect may be estimated from discrepancies between our mean profile and profiles measured by L15 within several days (less than one Carrington period), as illustrated in Figure~\ref{fig:obsEvolution}. They seem to be at a level of $\sim 15$\% for locations close to solar equator. At larger heliolatitudes, there is additional uncertainty related to the heliolatitude dependence of the total irradiance $I_{\mathrm{tot}}$, which currently cannot be verified experimentally, but is believed to be another $\sim 15$\% (comparable to the magnitude of the departure of the pole-to-equator ratio from 1).

An uncertainty that can be quantified is the influence of the uncertainty in the absolute calibration of $I_{\mathrm{tot}}$. This uncertainty is on the order of 15\% \citep{woods_etal:05a}. Since the mathematical form of our model (Equation~\ref{eq:modelFinal}) is strongly nonlinear in $I_{\mathrm{tot}}$, then if the $I_{\mathrm{tot}}$ time series used to find the coefficients of our model is biased by a factor $1 \pm \delta$, $0 \le \delta \ll 1$, then the entire model derivation procedure must be repeated with $I_{\mathrm{tot}}$ values rescaled by a $1 \pm \delta$, but the data also need to be identically rescaled. Therefore, the shapes of the profiles (both data and model) will not change, only their magnitudes will be modified by a factor $1 \pm \delta$.

We verified this by looking for differences between the radiation pressure models obtained as a result of the full fitting procedure with the bias factors $1 \pm \delta, \delta =  0.15$ applied to both $I_{\mathrm{tot}}$ and data, and the model resulting from simple rescaling of the nominal model, evaluated with the nominal (non-scaled) value of $I_{\mathrm{tot}}$, by the factor $1 \pm \delta$. We found no differences. Therefore, if a systematic rescaling of the $I_{\mathrm{tot}}$ by a factor of $1 \pm \delta$ is needed in the future, then the only change of the radiation pressure in our model will be multiplication by $1 \pm \delta$ and using the {\em old} (i.e., non-scaled) value of $I_{\mathrm{tot}}$. 

\section{Summary and conclusions}
\label{sec:summary}
Based on all available observations of the full-disk solar Lyman-$\alpha$ line (published by L15), performed using the SUMER instrument on board SOHO and covering almost a full cycle of solar activity, we found an analytical representation of the profile shape that is parametrized by the linewidth- and disk-integrated Lyman-$\alpha$ irradiance. The definition of the model is given in Equation~\ref{eq:modelFinal}. For a given value of the solar Lyman-$\alpha$ line-integrated irradiance and a given radial velocity $v_r$ of an atom, the model returns the radiation pressure coefficient $\mu(I_{\mathrm{tot}}, v_r)$ for hydrogen or deuterium atoms. The numerical values of the parameters to be used in Equation~\ref{eq:modelFinal} are listed in Table~\ref{tab:linCo} for H and in Table~\ref{tab:linCoD} for D. To calculate the $\mu$ factor for a given time $t$ it is recommended to use for $I_{\mathrm{tot}}$ the values obtained from linear approximation for $t$ between the neighboring Carrington averages of the daily time series of $I_{\mathrm{tot}}$ available from LASP \citet{woods_etal:05a}. To account for the hypothetical heliolatitudional dependence of $I_{\mathrm{tot}}$ and the resulting modification of the Lyman-$\alpha$ line profile it is recommended to use Equation~\ref{eq:helioLatiItot}.

The uncertainties of the present model are difficult to assess given all the complexities of the profile observations and observation deconvolution, the uncertainties related to the absolute scaling of the observed profiles due to uncertainties in the daily time series of $I_{\mathrm{tot}}$ of both random and systematic nature, and due to our fitting. We estimate, however, that the model is accurate to 15--20\%.

In comparison with the model of evolution of the solar Lyman-$\alpha$ line by TB09, the present model features larger horn-to-center ratios for both horns, especially for the conditions of a low solar activity. In fact, the present model suggests a close to linear dependence of these ratios on $I_\mathrm{tot}$ values, in agreement with observations. This will likely affect predictions of models of the density and flux of ISN H and D, of their derivative populations (PUIs, solar wind ENAs), and of the intensity of the helioglow. Survival probabilities of the ENAs observed at 1~AU are likely to be less affected. Detailed studies of these effects are underway and will be published in a separate paper in a near future.

\acknowledgments
This study was supported by Polish National Science Center grant 2015-18-M-ST9-00036.

\bibliographystyle{aasjournal}
\bibliography{iplbib}

\end{document}